\documentclass[english,3p,times,twocolumn,preprint,12pt]{elsarticle}
\usepackage[T1]{fontenc}
\usepackage[latin9]{inputenc}
\usepackage{amsmath}
\usepackage{amssymb}
\usepackage{graphicx}

\usepackage{babel}

\journal{Physics Letter A}
\begin{document}

\title{Quantum random walk in periodic potential on a line}

\author{Min Li}

\ead{mickeylm@mail.ustc.edu.cn}

\author[]{Yong-Sheng Zhang\corref{cor1}}

\ead{yshzhang@ustc.edu.cn (+86-0551-3607340)}

\author{Guang-Can Guo}

\cortext[cor1]{Corresponding author.}

\address{Key Laboratory of Quantum Information, University of Science and
Technology of China, CAS, Hefei, 230026, People's Republic of China}

\begin{abstract}
We investigated the discrete-time quantum random walks on a line in
periodic potential. The probability distribution with periodic potential
is more complex compared to the normal quantum walks, and the standard
deviation $\sigma$ has interesting behaviors for different period
$q$ and parameter $\theta$. We studied the behavior of standard
deviation with variation in walk steps, period, and $\theta$. The
standard deviation increases approximately linearly with $\theta$
and decreases with $1/q$ for $\theta\in(0,\pi/4)$, and increases
approximately linearly with $1/q$ for $\theta\in[\pi/4,\pi/2)$.
When $q=2$, the standard deviation is lazy for $\theta\in[\pi/4+n\pi,3\pi/4+n\pi],n\in Z$.\end{abstract}
\maketitle

\section{Introduction}

Quantum walks, as the quantum version of the classical random walks,
were first introduced in 1993 \citep{Aharonov}. Recently, quantum
walks have attracted great attention from mathematicians, computer
scientists, physicists, and engineers. (For an introduction, see Ref.
\citep{Nayak,Kempe}). Some new quantum algorithms based on quantum
walks have already been proposed \citep{Shenvi2003,Childs2004,Ambainis,Ambainis2004,Childs2003,Childs2002}.
They proved that a discrete time quantum walk can be used to perform
an oracle search on a database of $N$ items with $O(\sqrt{N})$ calls
to the oracle \citep{Shenvi2003}, and also can be used for universal
computation \citep{Childs2009,lovett}.

Quantum walks in many different situations have been studied extensively.
For example, the quantum walks in graph \citep{Aharonov-1}, on a
line with a moving boundary \citep{Kwek}, with multiple coins \citep{Brun-1}
or decoherent coins \citep{Brun}.

However, quantum walks in periodic potential has not been studied
yet. This kind of quantum walks are popular in physics. For example,
the motion of the atom in the double well lattice \citep{double-well}
and the propagation of photon in  periodically varying the coupling 
in waveguide lattice \citep{waveguide,optical-mesh-lattices}
with different waveguide periods or in the beam splitters array \citep{beam-splitter}
with two kinds of BS at periodic vertices. In this paper, we will
present the behaviors of quantum walks in periodic potential. We will
discuss the probability distribution and the standard deviation for
different periods, potentials and steps.

\section{Normal quantum wlaks and quantum scattering walks}

In this paper, we concern with the discrete-time quantum walks. To
be consistent, we adopt analogous definitions and notations as those
outlined in \citep{Travaglione}. The total Hilbert space is given
by $\mathcal{H}\equiv\mathcal{H}_{P}\otimes\mathcal{H}_{C}$, where
$\mathcal{H}_{P}$ is spanned by the orthonormal vectors $\left\{ \mid x\rangle\right\} $
which representing the position of the walker and $\mathcal{H}_{C}$
is the two-dimensional coin space spaned by two orthonormal vectors
which are denoted as $\mid\downarrow\rangle$ and $\mid\uparrow\rangle.$

Each step of the quantum walk can be split into two operations: the
flip of a coin and the position motion of the walker according to
the coin state.

Here, for simplicity, we choose a Hadamard coin as the normal quantum
walk's coin, so the coin operator can be written as 

\begin{equation}
\hat{H}\mid\downarrow\rangle=\frac{1}{\sqrt{2}}(\mid\downarrow\rangle+\mid\uparrow\rangle),\:\hat{H}=\frac{1}{\sqrt{2}}\left(\begin{array}{ccc}
1 &  & 1\\
1 &  & -1
\end{array}\right).
\end{equation}

The position displacement operator is given by

\begin{equation}
\hat{S}=e^{i\hat{p}\hat{\sigma}_{z}}=\sum_{x}\hat{S}_{x},
\end{equation}
where $\hat{p}$ is the momentum operator, $\hat{\sigma}_{z}$ is
the Pauli-$z$ operator, 

\begin{equation}
\hat{S}_{x}=\mid x+1\rangle\langle x\mid\otimes\mid\uparrow_{x}\rangle\langle\uparrow_{x}\mid+\mid x-1\rangle\langle x\mid\otimes\mid\downarrow_{x}\rangle\langle\downarrow_{x}\mid.
\end{equation}

Therefore, the state of the walker after $N$ steps is given by

\begin{equation}
\begin{aligned}\mid\Psi_{N}\rangle & =\left[\hat{S}(\hat{I}_{P}\otimes\hat{H}_{C})\right]^{N}\mid\Psi_{0}\rangle\\
 & =\left[\sum_{x}\hat{S}_{x}(\hat{I}_{P}\otimes\hat{H}_{C})\right]^{N}\mid\Psi_{0}\rangle,
\end{aligned}
\label{eq:hadamardDQW}
\end{equation}
where $\mid\Psi_{0}\rangle$ is the initial state of the system. 

The rule of quantum scattering walks we use here was described in
Ref. \citep{Feldman,Hillery}. Suppose that the state is in $\mid j+1,j\rangle,$
which means that in the last step the walker walked from $\mid j+1\rangle$
to $\mid j\rangle$. In the next step if it is transmitted, it will
be in the state $\mid j,j-1\rangle$, and if it is reflected it will
be in the state $\mid j,j+1\rangle$. Then we have the transition
rule

\begin{equation}
\hat{U}\mid j+1,j\rangle=t\mid j,j-1\rangle+r\mid j,j+1\rangle,\label{eq:scatter}
\end{equation}
where $t$ and $r$ are the transmission and reflection coefficients
respectively, the unitarity implies that $\mid t\mid^{2}+\mid r\mid^{2}=1$.

If we use $\left\{ \mid\downarrow\rangle,\mid\uparrow\rangle\right\} $
to represent the direction that the walker just walked, Eq. \eqref{eq:scatter}
can be written as:
\begin{equation}
\begin{aligned}\hat{U}\mid j,\downarrow\rangle & =t\mid j-1,\downarrow\rangle+r\mid j+1,\uparrow\rangle\\
 & \equiv\hat{S}_{j}\hat{C}\mid j,\downarrow\rangle.
\end{aligned}
\label{eq:scatterDQW}
\end{equation}

The unitarity of the scattering gives the transformation matrixes

\begin{equation}
C_{1}=\left(\begin{array}{ccc}
t &  & r^{*}\\
r &  & -t^{*}
\end{array}\right)\textrm{ or }C_{2}=\left(\begin{array}{ccc}
t &  & -r^{*}\\
r &  & t^{*}
\end{array}\right).\label{eq:c}
\end{equation}

From Eq. \eqref{eq:scatterDQW}, we can know that the quantum scattering
walk is the same as the coined quantum walk \citep{Hillery}. Without
loss of generality, we choose $t=\sin\theta$ and $r=\cos\theta$,
and use the form of $C_{1}$. Then the scattering matrix can be written
as 

\begin{equation}
\hat{C}=\left(\begin{array}{ccc}
\sin\theta &  & \cos\theta\\
\cos\theta &  & -\sin\theta
\end{array}\right).
\end{equation}

The case of $\theta=\pi/4$ corresponds to the discrete quantum walks
with Eq. \eqref{eq:hadamardDQW}.

\begin{figure}
\begin{centering}
\includegraphics[width=20pc]{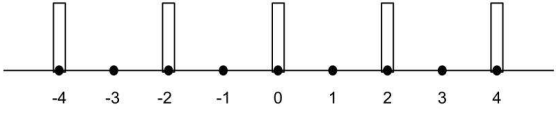}
\par\end{centering}

\caption{Periodic potential on a line, for period $q=2$. }
\label{fig:period}
\end{figure}

\section{Quantum walks in periodic potential}

We will consider the case that a normal quantum walker walks in periodic
potential like in Fig. \ref{fig:period}.

As a model, we consider the situation that the walker walks as scattering
quantum walk at the positions with potential, and walks as normal
discrete quantum walk for the rest. The potential is described by
parameter $\theta$. Then, the operator can be written as

\begin{equation}
U=\sum_{x=nq,n\in Z}\hat{S}_{x}\hat{C}+\sum_{x\neq nq,n\in Z}\hat{S}_{x}\hat{H},
\end{equation}
where $q$ is the period, $n$ is an integer, and the final state
after $N$ steps is given by

\begin{equation}
\mid\Psi_{N}\rangle=U^{N}\mid\Psi_{0}\rangle.
\end{equation}
Here the initial state we use is 

\begin{equation}
\mid\Psi_{0}\rangle=\frac{1}{\sqrt{2}}\mid0\rangle(\mid\downarrow\rangle+i\mid\uparrow\rangle).
\end{equation}

\begin{figure}
\begin{centering}
\includegraphics[width=3in]{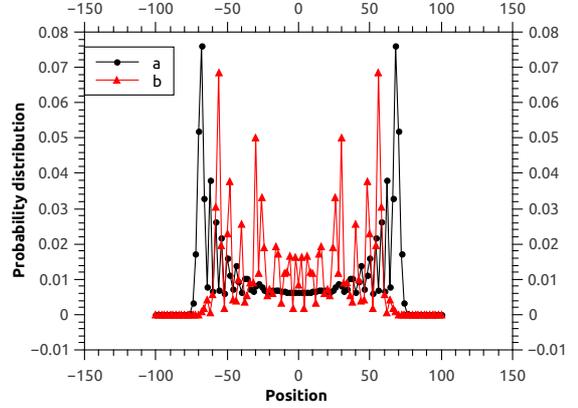}
\par\end{centering}

\caption{(Color online) Probability distribution for a normal quantum walk
(a) on a line after $100$ steps with the initial state $\frac{1}{\sqrt{2}}\mid0\rangle(\mid\downarrow\rangle+i\mid\uparrow\rangle)$
and a Hadamard coin, as well as for quantum walk but in periodic potential
(b) with the period $q=4$, $\theta=\pi/6$ and the same initial state.}
\label{fig:pro-distribution}
\end{figure}

Fig. \ref{fig:pro-distribution} shows the probability distribution
after $N=100$ steps of the quantum walk starting from $\mid\Psi_{0}\rangle$
with and without periodic potential. In the first, we notice that
the existence of the periodic potential does not change the symmetric
of probability distribution. In the second, the behavior of probability
distribution of quantum walks in periodic potential is more complex
than normal quantum walks.

\begin{figure}
\centering{}\includegraphics[width=3in]{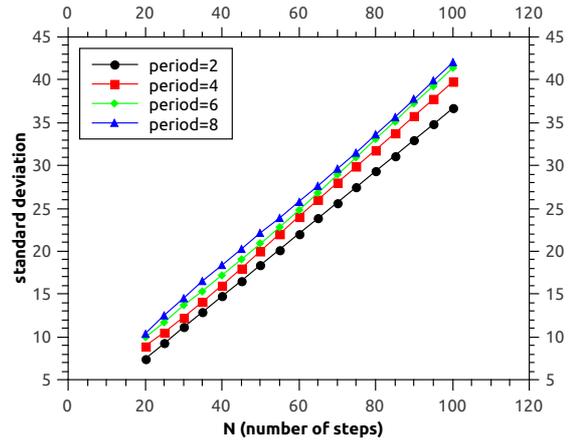}\caption{(Color online) Standard devitation $\sigma$ for quantum walk in periodic
potential with different periods when $\theta=\pi/6$.}
\label{fig:pi6-step-st-angle}
\end{figure}

\begin{figure}
\centering{}\includegraphics[width=3in]{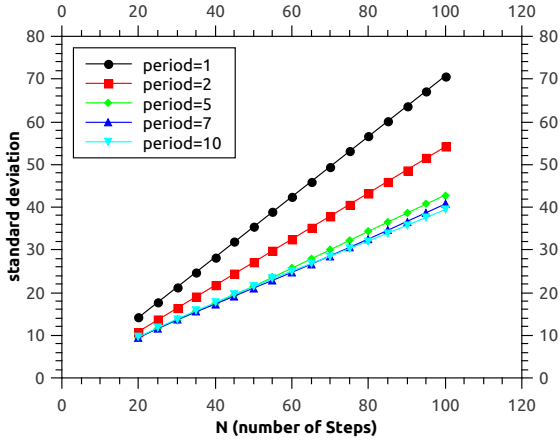}\caption{(Color online) Standard devitation $\sigma$ for quantum walk in periodic
potential with different periods when $\theta=\pi/3$.}
\label{fig:pi3-step-st-angle}
\end{figure}

Fig. \ref{fig:pi6-step-st-angle} and Fig. \ref{fig:pi3-step-st-angle}
show the standard deviation $\sigma=\sqrt{\left\langle (x-\left\langle x\right\rangle )^{2}\right\rangle }$
for quantum walks in periodic potential with different periods when
$\theta=\pi/6$ and $\theta=\pi/3$ respectively. Firstly, we can
know that, regardless of the existence of period potential with different
period $q$ and $\theta$, the standard deviation still increases
approximately linearly with $N$ (number of steps). Secondly, when
$\theta=\pi/6$, the standard deviation increases with the period
increasing (Fig. \ref{fig:pi6-step-st-angle}), but if $\theta=\pi/3$
the standard deviation decreases with the period increasing (Fig.
\ref{fig:pi3-step-st-angle}).

\begin{figure}
\begin{centering}
\includegraphics[width=3in]{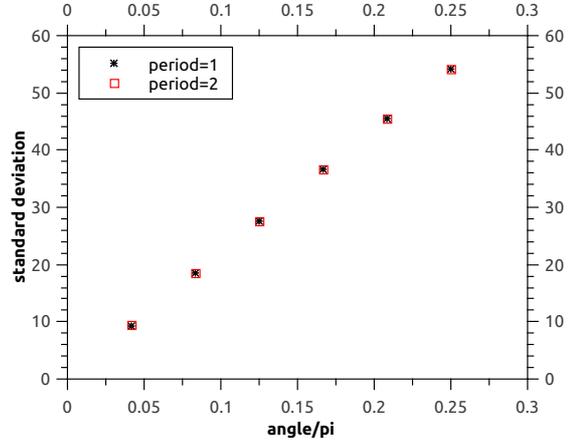}
\par\end{centering}

\caption{(Color online) Standard deviation $\sigma$ for different $\theta\in(0,\pi/4),$
with period $q=1$ (black star) and 2 (red rectangle), after 200 steps
of quantum walk.}
\label{fig:period-1-2}
\end{figure}

Fig. \ref{fig:period-1-2} shows the standard deviation $\sigma$
for period $q=1\text{ and }2$, with different $\theta\in(0,\pi/4)$.
Then we can know that the standard deviations are nearly the same
when $\theta\in(0,\pi/4)$.

\begin{figure}
\begin{centering}
\includegraphics[width=3in]{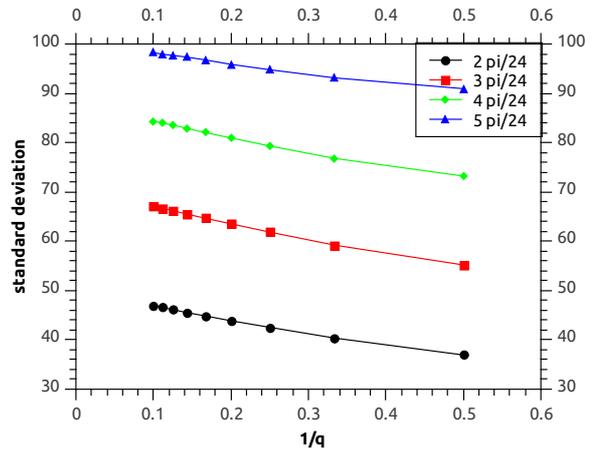}
\par\end{centering}

\caption{(Color online) Standard deviation $\sigma$ for (1/period) with different
$\theta\in(0,\pi/4),$ after 200 steps of quantum walk.}
\label{fig:period-st-0-pi4}
\end{figure}

\begin{figure}
\begin{centering}
\includegraphics[width=3in]{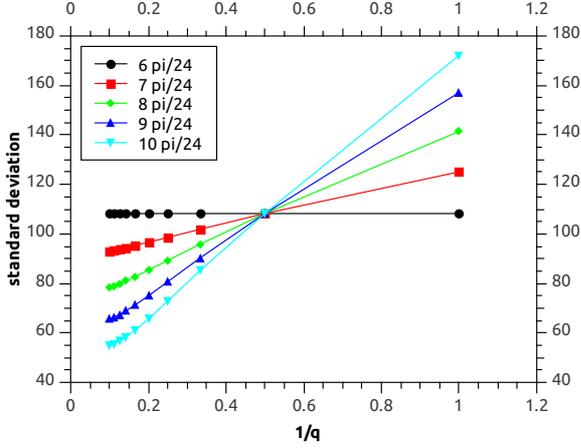}
\par\end{centering}

\caption{(Color online) Standard deviation $\sigma$ for (1/period) with different
$\theta\in[\pi/4,\pi/2),$ after 200 steps of quantum walk.}
\label{fig:period-st-pi4-pi2}
\end{figure}

Fig. \ref{fig:period-st-0-pi4} and Fig. \ref{fig:period-st-pi4-pi2}
show the standard deviation $\sigma$ for different periods with different
$\theta\in(0,\pi/2)$, after 200 steps of quantum walk. From the figures,
we can know the standard deviation decreases approximately linearly
with $1/q$ when $\theta\in(0,\pi/4)$, $2\leq q\leq10$ (Fig. \ref{fig:period-st-0-pi4}),
and increases approximately linearly with $1/q$ when $\theta\in[\pi/4,\pi/2)$,
$1\leq q\leq10$ (Fig. \ref{fig:period-st-pi4-pi2}).

\begin{figure}
\begin{centering}
\includegraphics[width=3in]{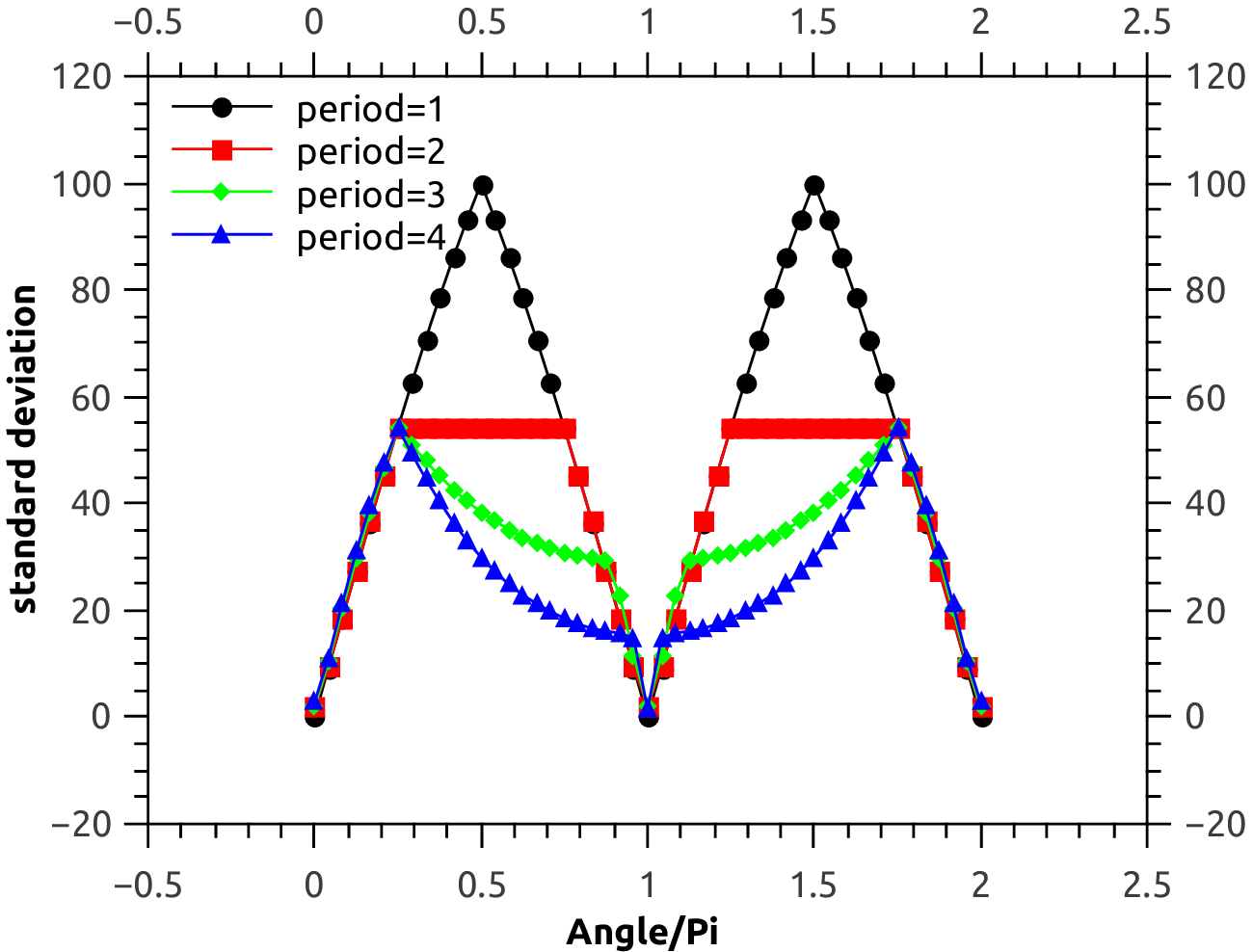}
\includegraphics[width=3in]{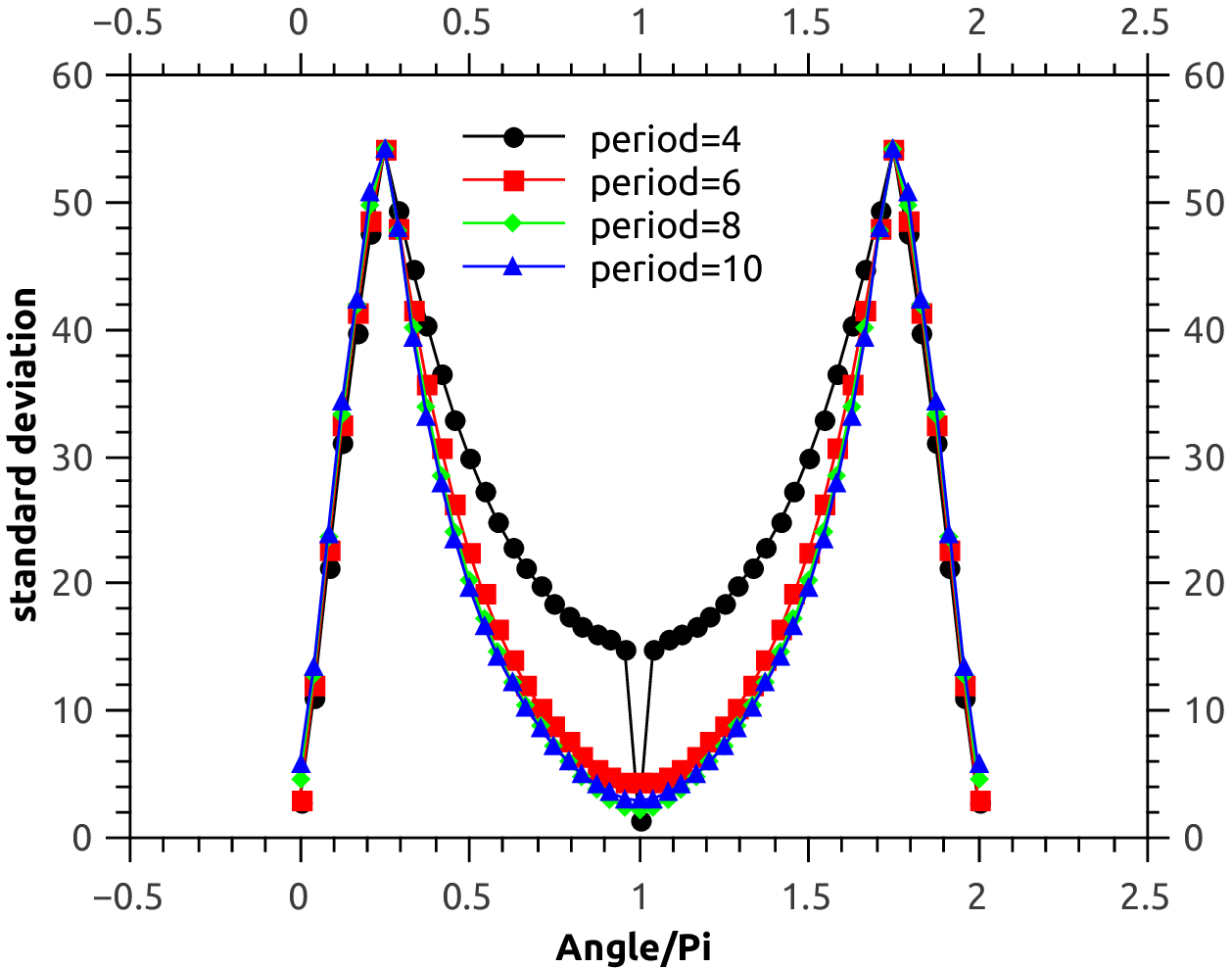}
\par\end{centering}

\caption{(Color online) Standard deviation $\sigma$ for quantum walker walk
100 steps in periodic potential with different period and $\theta$.}
\label{fig:period-angle-st}
\end{figure}

\begin{figure}
\begin{centering}
\includegraphics[width=3in]{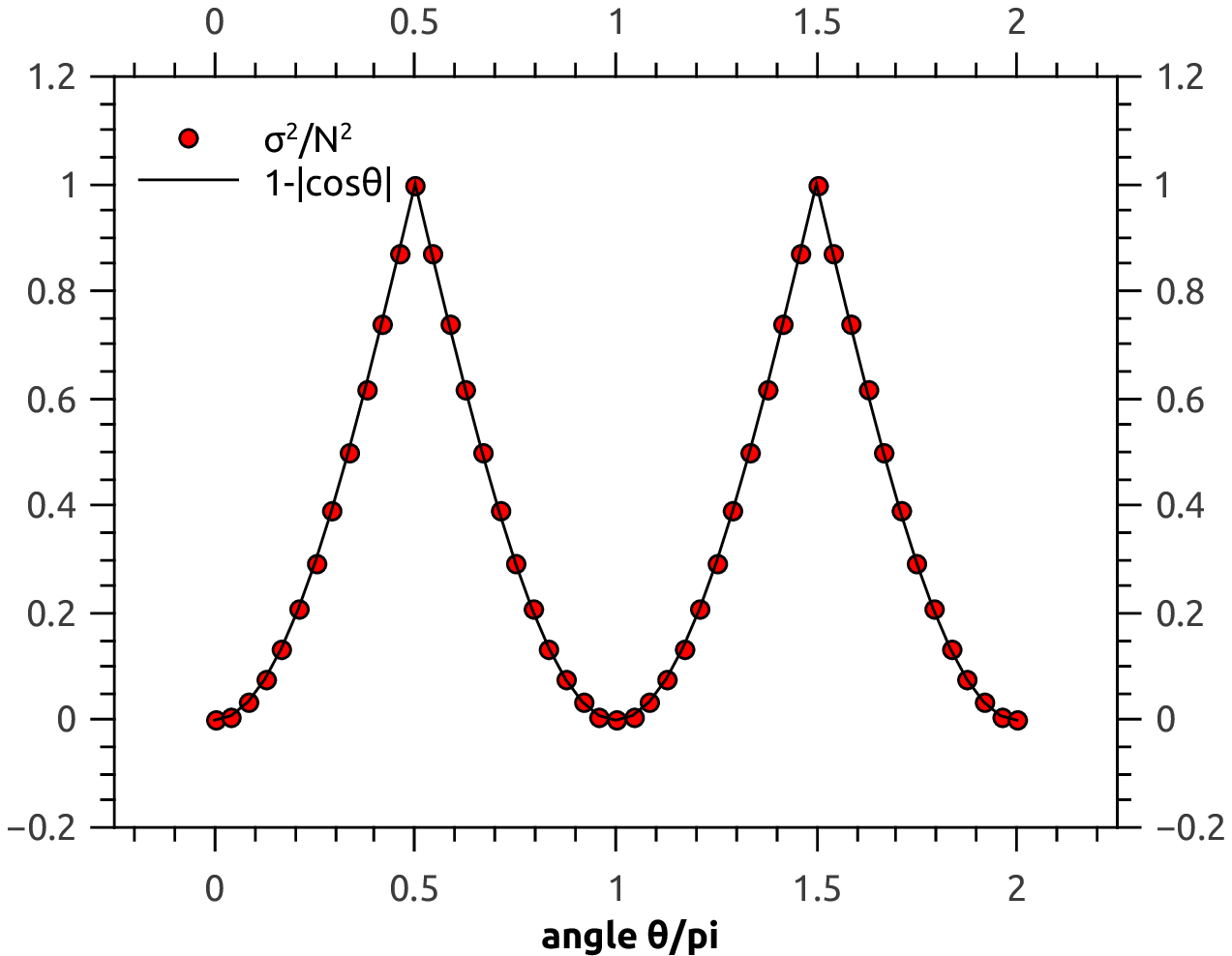}
\par\end{centering}

\caption{(Color online) Variation of $\sigma^{2}/N^{2}$ for quantum walks
in periodic potential with $q=1$, where $\sigma$ is the standard
deviation and $N$ is the number of steps, and the function $f(\theta)=1-|\cos\theta|$
.}
\label{fig:q1-1-cos}
\end{figure}

Fig. \ref{fig:period-angle-st} shows the standard deviation for the
quantum walks in periodic potential with $\theta$ from $0$ to $2\pi$
and different periods. We can know that when $\theta\in[0,\pi/4]$,
standard deviation increases approximately linearly with $\theta$
for different period. For all periods, the line $\theta=\pi$ is a
symmetry axis, so $\sigma(\theta)=\sigma(2\pi-\theta).$

When the period $q>3$, and $\theta\in[\pi/4,\pi]$, the standard
deviation decreases with the increasing of $\theta$. The case of
period $q=2$, is the transition state between $q=1$ and $q=3$.
The standard deviation is nearly the same as $\theta\in[\pi/4,3\pi/4]$,
while the transmission coefficient $t=\sin\theta$ is larger than
the reflection coefficient $r=\cos\theta$.

When the period $q=1$, i.e., there is no Hadamard walk but the scattering
walk. With the increasing of $\theta$ from $0$ to $\pi/2$, the
standard deviation will increase approximately linearly \citep{Chandrashekar},
and from $\pi/2$ to $\pi$, it will decrease approximately linearly.
The standard deviation function has a period of $\pi$. With the increasing
of $\theta\in[0,\pi/2]$, the transmission coefficient $t=\sin\theta$
increases and reflection coefficient $r=\cos\theta$ decreases, then
the diffusion velocity increases, so the standard deviation increases.
Conversely, $t$ decreases and $r$ increases while $\theta$ increases
from $\pi/2$ to $\pi$, then the standard deviation decreases with
the increasing of $\theta\in[\pi/2,\pi]$. The case that $\theta\in[\pi,2\pi]$
is the same as $\theta\in[0,\pi]$. From Ref. \citep{Chandrashekar},
we can know
\begin{equation}
\sigma^{2}\approx(1-\sin\theta')N^{2}=(1-\cos\theta)N^{2},
\end{equation}
where $\theta'$ has been defined in Ref. \citep{Chandrashekar} that
is equals to $\pi/2-\theta$, when $\theta'\in[0,\pi/2].$ 

Fig. \ref{fig:q1-1-cos} shows $\sigma^{2}/N^{2}$ as a function of
$\theta$, and the function $f(\theta)=1-|\cos\theta|$ . From the
figure, we can know that 
\begin{equation}
\sigma\approx\sqrt{1-|\cos\theta|}\, N,\label{eq:1-cos}
\end{equation}
 for any $\theta$, when $q=1$. Eq. \eqref{eq:1-cos} is more precise
than linear approximation $\sigma\approx\begin{cases}
(2\theta/\pi)N, & \theta\in[2n\pi,(2n+1)\pi]\\
(2-2\theta/\pi)N, & \theta\in[(2n-1)\pi,2n\pi]
\end{cases},n\in Z$.

\section{Conclusion}

In this paper, we have discussed the one-dimensional quantum walks
in the presence of a periodic potential. The behavior of probability
distribution, standard deviation of quantum walks in periodic potential
is different from the quantum walks without periodic potential. When
$\theta=\pi/4$, the quantum walks with periodic potential are the
same as the normal quantum walks. The case that period of the potential
$q=1$ and $\theta=\pi/2$ corresponds to free particle propagation,
then after $t$ steps, $\mid\Psi(t)\rangle=1/\sqrt{2}(\mid(-t)L\rangle+i\mid tR\rangle$.
If $\theta=n\pi,n\in\mathbb{Z}$, the transmission probability decreases
to $0$, and the reflection probability is $1$, this situation implies
the walker walks between two infinite high walls which are $q$ apart,
and this situation is similar to quantum walks in cycle graph. Next,
we can know that the standard deviation increases approximately linearly
with $\theta$ and decreases with $1/q$ if $\theta\in(0,\pi/4)$,
and increases approximately linearly with $1/q$ if $\theta\in[\pi/4,\pi/2)$.
When $q=2$, the quantum walk is lazy for $\theta\in[\pi/4+n\pi,3\pi/4+n\pi],n\in Z$,
while the transmission coefficient is larger than the reflection coefficient.
Then we can know some property about quantum walks in double well
optical lattice \citep{double-well} and perodic waveguide lattice
\citep{waveguide} consists of different waveguids.

\section*{Acknowledgments}

This work was supported by the National Natural Science Foundation
of China (Grant No. 10974192, 61275122), the National Fundamental
Research Program of China (Grant No. 2011CB921200, 2011CBA00200),
and K. C. Wong Education Foundation.


\end{document}